\documentclass[%
 reprint,
 superscriptaddress,
 amsmath,amssymb,
 aps,
 longbibliography
]{revtex4-1}

\usepackage{graphicx}% Include figure files
\usepackage{dcolumn}% Align table columns on decimal point
\usepackage{bm}% bold math
\usepackage{url}
\usepackage{xcolor}

\begin{document}

\preprint{APS/123-QED} \title{Statistical analysis of electronic and phononic
  transport simulations of metallic atomic contacts}

\author{D. O. M\"{o}hrle}
 \affiliation{Department of Physics, University of Konstanz, 78457 Konstanz, Germany}
 \affiliation{German Aerospace Center, Pfaffenwaldring 38-40, 70569 Stuttgart, Germany}
\author{F. M\"{u}ller}
 \affiliation{Department of Physics, University of Konstanz, 78457 Konstanz, Germany}
\author{M. Matt}
 \affiliation{Department of Physics, University of Konstanz, 78457 Konstanz, Germany}
\author{P. Nielaba}
 \affiliation{Department of Physics, University of Konstanz, 78457 Konstanz, Germany}%Lines break automatically or can be 
\author{F. Pauly}
 \affiliation{Okinawa Institute of Science and Technology Graduate University, Onna-son, Okinawa 904-0495, Japan}
 \affiliation{Department of Physics, University of Konstanz, 78457 Konstanz, Germany}%Lines break automatically or can be 

\date{\today}% It is always \today, today,
             %  but any date may be explicitly specified
             
\begin{abstract}

We adapt existing phonon heat transport methods to compute the phononic
thermal conductance of metallic atomic contacts during a stretching process.
Nonequilibrium molecular dynamics simulations are used to generate
atomic configurations and to simultaneously determine the phononic thermal
conductance. Combining the approach with established electronic structure
calculations based on a tight-binding parametrization allows us to calculate
in addition charge transport properties of each contact geometry within the
Landauer-B\"uttiker formalism. The method is computationally fast enough to
perform a statistical analysis of many stretching events, and we apply it here
to atomic junctions formed from three different metals, namely gold (Au),
platinum (Pt) and aluminum (Al). The description of both phononic and
electronic contributions to heat transport allows us to examine the validity
of the Wiedemann-Franz law at the atomic scale. We find that it is well obeyed
in the contact regime at room temperature for Au and Al as far as only
electronic contributions are concerned, but deviations of up to 10\% arise for
Pt. If the total thermal conductance is studied, deviations of typically less
than 10\% arise for Au and Al, which can be traced back mainly to phononic
contributions to the thermal conductance, while electronic and phononic
contributions can add up to some 20\% for single-atom contacts of Pt.

%\begin{description}
%\item[Usage]
%Secondary publications and information retrieval purposes.
%\item[PACS numbers]
%May be entered using the \verb+\pacs{#1}+ command.
%\item[Structure]
%You may use the \texttt{description} environment to structure your abstract;
%use the optional argument of the \verb+\item+ command to give the category of each item. 
%\end{description}

\end{abstract}

\pacs{Valid PACS appear here}% PACS, the Physics and Astronomy
                             % Classification Scheme.
%\keywords{Suggested keywords}%Use showkeys class option if keyword
                              %display desired
\maketitle

%\tableofcontents

\section{\label{sec:introduction}Introduction}

The ongoing miniaturization of electronic devices promises an increase in
speed and efficiency of operation. The physical limit of this
miniaturization is given by integrated circuits consisting of functional units
and wires made of single atoms. However, at the nanoscale heat dissipation
becomes an increasingly important issue \cite{pop2010}, and thermal management
is of high relevance.

In this context, metallic atomic contacts serve as ideal systems to test
charge and energy transport at the atomic scale
\cite{agrait2003,cuevas2017}. Crucial progress in understanding thermopower
\cite{ludoph1999,evangeli2015}, heat dissipation \cite{lee2013}, and heat
transport \cite{cui2017,mosso2017,mosso2019} in these systems has been made in
recent years. The advances enabled tests of the Wiedemann-Franz law at the
atomic scale, where transport is assumed to be phase coherent and elastic. In
contrast the Wiedemann-Franz law, which connects electrical and thermal
conductances, has been discovered empirically for macroscopic systems, where
incoherent and inelastic scattering events also play a role. For metallic
atomic contacts of Au and Pt only slight deviations from it were detected at
room temperature \cite{cui2017}.

Experimentally atomic contacts are realized using the scanning tunneling
microscope (STM) technique \cite{gimzewski1987, agrait1993, cui2017,
  mosso2017}, in which a STM tip is repeatedly dipped into a substrate and
slowly separated again, or the mechanically controllable break junction method
\cite{muller1992, krans1993, scheer1998}, which consists of a wire on a
flexible sample that is bent until it narrows to cross sections of single
atoms and finally breaks. A drawback of both methods is that in the contact
regime, positions of individual atoms are not under precise control. For this
reason different traces of electrical and thermal conductance versus distance
appear, and data is analyzed statistically in terms of histograms to identify
physical properties \cite{agrait2003,cui2017,mosso2017,mosso2019}.

To understand the transport properties of atomic contacts, theoretical methods
are typically applied to single junction geometries, which are assumed to
correspond to an experimentally relevant geometry
\cite{agrait2003,cuevas2017}. Theoretical work that generates a large amount
of charge or energy transport data and analyzes it statistically is still
scarce
\cite{dreher2005,pauly2006,pauly2011,makk2011,chen2014,evangeli2015,vardimon2016,kloeckner2017}.

Recent theoretical work found the phononic contribution $\kappa_{\text{ph}}$
to the total thermal conductance $\kappa$ to be small (less than 10\%) in
metallic atomic contacts of the heavy metals Au and Pt \cite{kloeckner2017}.
For the light Al, contributions of up to 20\% were predicted due to the higher
Debye energy $E_{\text{D}}$ as compared to Au \cite{kloeckner2017}, which
increases $\kappa_{\text{ph}}$. For a particular single-atom junction
geometry, the ratio $\kappa_{\text{ph}}/\kappa$ has even been computed to
reach as much as 40\% \cite{kloeckner2017}, which was mainly due to a
suppression of the electronic transmission and hence the electronic thermal
conductance $\kappa_{\text{el}}$.

Here we extend our previous work \cite{kloeckner2017} in order to study
electronic and phononic contributions to the thermal conductance for a
statistically significant set of atomic junction geometries. For this purpose
we combine a nonequilibrium molecular dynamics (NEMD) approach for the determination 
of junction geometries and of
phonon heat transport properties with a tight-binding method for the
electronic structure description and subsequent electronic transport
calculations \cite{cohen1994, mehl1996}. As compared to previous
  molecular-dynamics-based work \cite{dreher2005, pauly2006}, we are thus
  adding here the description of $\kappa_{\text{ph}}$. The combined approach
is computationally fast enough to enable the study of a larger number of
junction stretching processes. In this way, we are not only able to analyze
histograms of the electrical conductance and electronic thermal conductance
but also of the phononic heat conductance.

We choose the metals Au, Pt, and Al due to their different properties. It is
well known by now that the electrical conductance $G$ is quantized in
atomically thin gold nanocontacts \cite{costa1997, muller1996}, as a result of
$s$ orbitals being the main contributors to the transmission
\cite{scheer1998,dreher2005}.  For platinum, the $d$ orbitals also need to be
considered, leading to a larger conductance than for gold for similar atomic
configurations and to no conductance quantization \cite{muller1992,
  pauly2006}.  In aluminum, the $p$ orbitals show contributions in addition to
the $s$ orbitals \cite{scheer1997, pauly2008}. The electrical transport
properties are by now well understood in contrast to the thermal transport
properties, which became experimentally accessible at room temperature
only recently \cite{cui2017,mosso2017,mosso2019}. On the side of heat
transport, the different electronic properties of Au, Pt and Al are expected
to be reflected in the electronic heat conductance. Concerning phonon heat
transport the choice of the metals is also interesting, since the Debye
energies of Au and Pt of 15 and 20~meV, respectively, are very similar as well
as their atomic mass, while that of Al of 34~meV is nearly double the size
and the atoms are lighter by a factor of more than 7 \cite{ashcroft1976}.

In nanojunctions heat transport via photons may yield another contribution,
beside those of electrons and phonons discussed here. Indeed evanescent waves
can enhance radiative thermal transport contributions in nanogaps, as studied
in recent experiments \cite{kim2015, kloppstech2017, cui2017_2}. Since we
focus in this work on the contact regime and the precise magnitude depends on
assumptions about the macroscopic shape of the electrodes
\cite{kloeckner2017b}, we will neglect radiative heat transport. Estimates
yield some 10~pW/K for the photonic thermal conductance
\cite{cui2017,kloeckner2017b}, which is below the phononic thermal conductance
values obtained here and may amount to some 1\% of $\kappa$ for typical
junction geometries.

The paper is organized as follows. In Sec.~\ref{sec:theoretical_model} the
theoretical models used to generate junction geometries and to describe phonon
and electron heat transport are presented together with the Wiedemann-Franz
law.  Section~\ref{sec:results} discusses the results. For each of the three
materials, Au, Pt and Al, we show a single stretching process first, followed
by a statistical analysis of properties collected from $60$ such processes.
Finally, we summarize our conclusions in Sec.~\ref{sec:conclusions}. Some
  technical details, namely a full junction stretching process and the
  computational costs of our simulations, are presented in
  Sec.~\ref{sec:FullPull-ComputationalCost} of the appendix.

\section{\label{sec:theoretical_model}Theoretical model}

We use NEMD simulations to determine atomic configurations and phononic
thermal conductance values. Utilizing a tight-binding model, we compute the
electronic structure of each junction geometry and derive electronic transport
properties from it within the Landauer-B{\"u}ttiker formalism, namely the
electrical conductance and the electronic thermal conductance. We describe the
details of these procedures in the following.

\subsection{\label{subsec:Geometries}Junction geometries}

\begin{figure}
  \includegraphics[width=0.8\columnwidth]{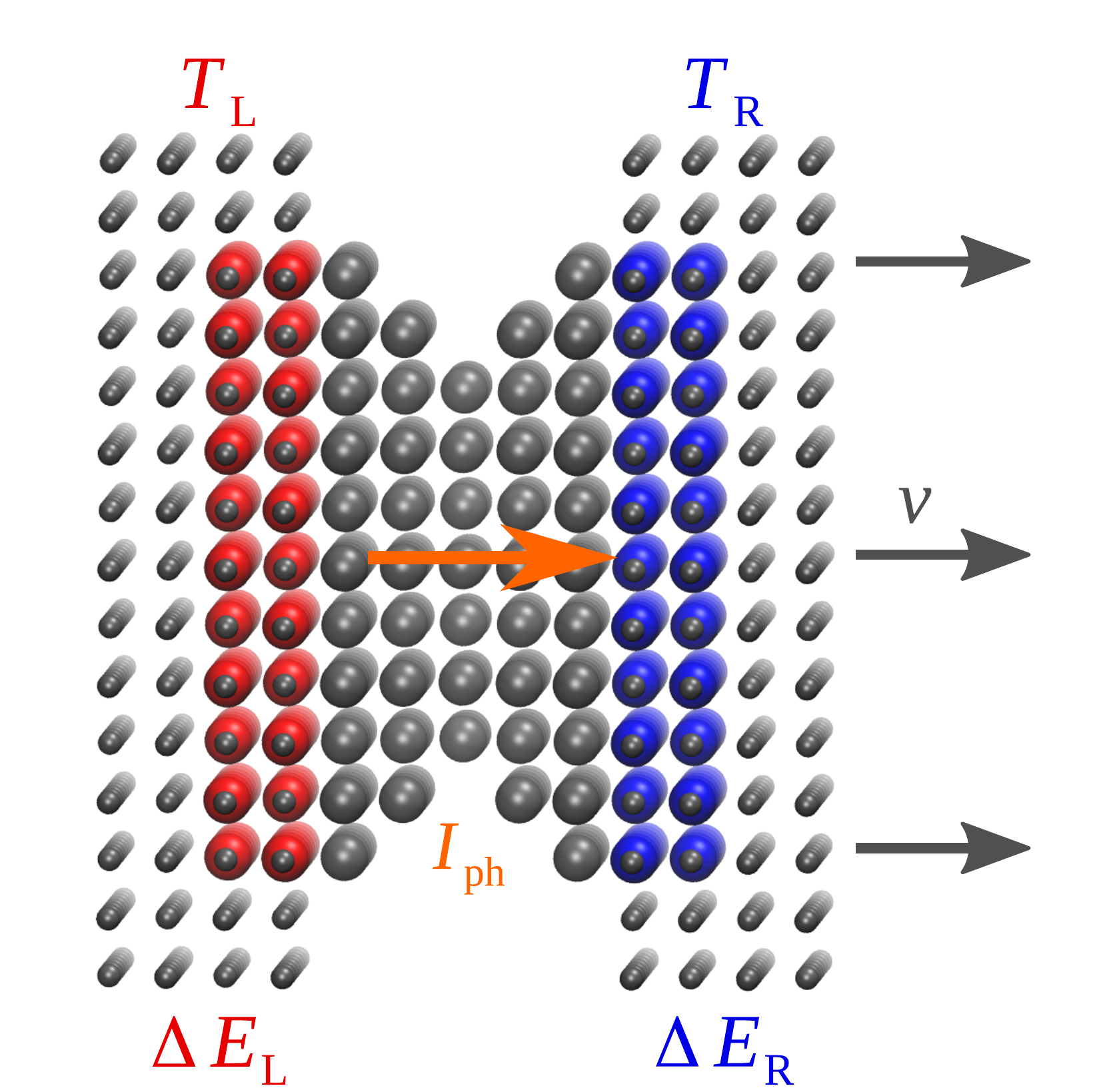}
  \caption{Starting configuration for the NEMD simulations. The transport in
    the square-shaped nanowire is oriented along the $(100)$ crystallographic
    direction. Interatomic potentials and lattice parameters are adjusted for
    the different metals. Two thermostats are applied to the red and blue
    reservoir atoms on the left (L) and right (R) side at actual temperatures
    $T_{\text{L}}$ and $T_{\text{R}}$. The energy changes $\Delta
      E_{\text{L}}$ and $\Delta E_{\text{R}}$ due to the thermostats are
      related to the heat transported by the phononic current $I_{\text{ph}}$,
      visualized by the orange arrow in the middle of the nanowire. Small
    atoms surrounding the thermal reservoir atoms are held fixed. During
      the simulation, the ``fixed'' atoms on the right change their positions
      at a constant velocity $v$, as indicated by three arrows. The large
    gray atoms between the thermal reservoirs are the wire atoms, whose motion
    is free of external constraints. We note that when trying to estimate
      the number of atoms in the junction from the picture, the face-centered
      cubic lattice geometry needs to be taken into account. In the side view
      shown not all of the atoms that appear to be in a front row are actually
      located in the same plane.}
  \label{fig:md_geometry}
\end{figure}

To determine junction geometries, we use NEMD simulations and the program
package LAMMPS \cite{plimpton1995, lammps}.  The forces between atoms are
parameterized by the embedded-atom method (EAM) \cite{finnis1984}, and the
potentials are taken from Ref.~\cite{sheng2011}.

The initial metallic atomic junction configurations are shown in
Fig.~\ref{fig:md_geometry}. These nanowires with a square-like cross section
are oriented along the $(100)$ crystallographic direction and consist of 1126
atoms in total. We have selected the $(100)$ direction, because studies of
  gold nanowires with a transmission electron microscope have revealed that
  just before rupture they are crystalline and display only atomic
  configurations, where $(100)$, $(110)$ or $(111)$ directions lie parallel to
  the stretching direction \cite{Rodrigues2000}. The $121$ large red and
$121$ large blue spheres denote atoms of the hot and cold reservoirs for
phonon heat transport simulations, respectively. The thermal reservoirs
  are connected via $226$ flexible wire atoms, represented by the large gray
  spheres. The five atomic layers of the wire are narrowing down stepwise
  towards the center. Small gray atoms surrounding the reservoirs have fixed
positions and interact with other atoms via their potential. As depicted in
Fig.~\ref{fig:md_geometry}, the reservoirs and surrounding fixed atom groups
are two atomic layers deep. The purpose of the fixed atoms, totaling a number
of $2\times 329$, is to stabilize the geometries of the thermal reservoirs and
in this way mimic the behavior of two semi-infinite crystalline
  electrodes. We have varied the size of the thermal reservoirs and found no
significant changes in the resulting phononic thermal transport.  Relatively
small reservoir sizes are thus chosen in order to decrease computation times,
especially for the electronic structure calculations to be described later.

The equations of motion of the wire and reservoir atoms are solved using the
Velocity-Verlet integration method with a time step of $1$~fs. Each of the two
reservoirs is kept at a nearly constant temperature of $T_{\text{L}} = 330$~K
and $T_{\text{R}} = 300$~K by a separate velocity rescale thermostat. The
velocity rescale thermostats, as implemented in LAMMPS, are used because they
provide a stable temperature for different reservoir and wire sizes and are
computationally efficient.

To simulate a junction stretching event, we start as follows. In the thermal
reservoirs and the wire part velocities of the atoms are selected at random in
such a way that a Gaussian distribution centered at $300$~K is
achieved. Accordingly both thermostats are set to a temperature of $300$~K.
After an equilibration time of $1$~ns thermostat L slowly increases the
temperature of its reservoir from $300$ to $330$~K over a duration of $1$~ns,
while thermostat R remains at the temperature of $300$~K. After an additional
equilibration time of $1$~ns the stretching process begins by separating the
group of fixed atoms surrounding the cold reservoir R at a constant velocity
of $v = 0.01$~m/s from the hot reservoir L. This leads to a narrowing of the
wire and a typical conductance-displacement trace comparable to those observed
in experimental setups.

\subsection{Phonon transport}

The thermostats used in the NEMD simulation rescale the velocities of the
$N_X$ atoms considered in reservoir $X=\text{L,R}$, if the difference between
the actual and the desired temperatures according to the equipartition theorem
is greater than the predefined \textit{window} setting of the thermostat:
\begin{equation}
  \langle E_X \rangle = \left\langle \sum\limits_{i=1}^{N_X} \frac{m_{i}}{2} v_{i}^{2} \right\rangle = \frac{3}{2} N_X k_{\text{B}} T_X.
\end{equation}
In the expression $E_X$ and $T_X$ are the actual energy and temperature of the
reservoir $X$, $v_i$ and $m_i$ are the velocity and mass of the $i$th particle
and $k_{\text{B}}$ is the Boltzmann constant. The rescaling leads to an energy
change,
\begin{equation}
  \Delta E_X = \alpha \langle E_X \rangle \left( 1 - \frac{T_{\text{th,X}}}{T_X} \right),\label{eqn:dE}
\end{equation}
where $T_X$ is the temperature of the system before the rescaling and
$T_{\text{th},X}$ is the target temperature of the thermostat. Due to the
\textit{fraction} factor $\alpha$, the thermostat does not rescale fully to
the desired temperature. (We choose the settings of the thermostat as
\textit{window}$=0.01$~K and \textit{fraction} $\alpha=0.8$ to allow small
fluctuations of the temperature.) Note that the sign of the energy change in
Eq.~(\ref{eqn:dE}) is positive, if the system needs to be cooled, and negative
if it needs to be heated. This choice of sign is consistent with the data
output of the thermostats in LAMMPS.

As visible in Fig.~\ref{fig:md_geometry}, the energy difference $-\Delta
E_{\text{L}}$ is supplied to the hot reservoir by thermostat L and $-\Delta
E_{\text{R}}$ to the cold reservoir by thermostat R during each timestep.  The
orange arrow indicates the phononic heat current $I_{\text{ph}}$, which
transports energy from the hot (L, red) to the cold (R, blue) reservoir. We
can thus determine the phononic thermal conductance using Fourier's law via
\begin{equation}
  \kappa_{\text{ph}} = \frac{I_{\text{ph}}}{T_{\text{L}} - T_{\text{R}}} = \frac{1}{2\Delta t} \frac{-\Delta E_{\text{L}} + \Delta E_{\text{R}}}{T_{\text{L}} - T_{\text{R}}}. \label{eqn:kappa-ph-compute}
\end{equation}
We store the energy changes $\Delta E_{\text{L}},\Delta E_{\text{R}}$ and
temperatures $T_{\text{L}},T_{\text{R}}$ every $1000$ simulation time steps,
corresponding to $\Delta t=1$~ps. Since the fluctuations of $\Delta
E_{\text{L}}$ and $\Delta E_{\text{R}}$ are large, we smoothed $\Delta
E_{\text{L}},\Delta E_{\text{R}},T_{\text{L}},T_{\text{R}}$ by using a moving
average with a Gaussian weight function, cut off at each side at the standard
deviation of $1$~ns. From this data we compute $\kappa_{\text{ph}}$ via
Eq.~(\ref{eqn:kappa-ph-compute}). In order to compress the data to the 1~ns
timestep used in the electronic transport calculations to be described next,
we performed an arithmetic average over $1$~ns around the time point of the
prospective electronic transport calculation to obtain the final value for
$\kappa_{\text{ph}}$.  Due to the classical NEMD simulations we expect
  that studies of phonon heat flow with the present technique are limited to
  temperatures above 100~K.

\subsection{Electron transport\label{subsec:electron-transport}}

To calculate the electronic transport properties, we use the
Landauer-B\"uttiker scheme.  For a detailed description we refer to
Refs.~\cite{dreher2005,pauly2006,cuevas2017,kloeckner2017}.

In the linear-response regime the Landauer-B\"uttiker formalism states that
the electrical conductance is given by
\begin{equation}
	G = G_{0} K_0
\end{equation}
and the electronic thermal conductance by
\begin{equation}
  \kappa_{\text{el}} = \frac{2}{hT} \left( K_2 - \frac{K_1^2}{K_0} \right)
\end{equation}
with
\begin{equation}
  K_n = \int dE \left( E-\mu \right)^n \tau(E) \left( \frac{-\partial
    f(E,T)}{\partial E} \right),
\end{equation}
the conductance quantum $G_{0} = 2e^{2}/h$, the electronic transmission
$\tau(E)$, and the Fermi function $f(E,T)$. In our calculations, we will set
the electronic temperature to the average of the target temperatures of the
two thermostats, $T=315$~K.

We need to have information about the electronic structure in order to compute
the electronic transmission $\tau(E)$. For this purpose we employ a
tight-binding model \cite{cohen1994, mehl1996}, which parameterizes the
Hamiltonian through the Slater-Koster approximation \cite{slater1954}. Since
the bulk behavior differs from those at surfaces or in nanowires, we modified
the approach to ensure self-consistently a local charge neutrality of the wire
atoms, as explained in previous work \cite{dreher2005,pauly2006}.

For the calculation of the electronic transmission and hence $G$ and
$\kappa_{\text{el}}$ we redefine the central ``wire'' part as compared to the
phonon heat transport calculations discussed before. For every geometric
configuration the central part is chosen to consist of the two thermal
reservoirs (red and blue large spheres) as well as the large gray wire atoms,
totaling $468$ atoms. The two outer layers of the fixed atoms on both
  sides (small gray spheres in Fig. \ref{fig:md_geometry}, $225$ atoms on each
  side) are assumed to belong to semi-infinite, perfect face-centered cubic
  electrode crystals. The 104 fixed atoms on each side, surrounding the
colored phonon thermal reservoirs perpendicular to the heat flow, are ignored
for the electronic calculations to shorten computation times.

\subsection{\label{subsec:wiedemannfranz}Wiedemann-Franz law}

The empirical Wiedemann-Franz law connects the electrical conductance $G$ with
the electronic thermal conductance $\kappa_{\text{el}}$ via
\begin{equation}
  \kappa_{\text{el}} = L_{0} T G.\label{eqn:WFlaw}
\end{equation}
Here the Lorenz number is given by $L_{0} = (\pi^{2}/3) (k_{\text{B}}/e)^{2}$
with the elementary electronic charge $e=|e|$.

Equation~(\ref{eqn:WFlaw}) allows one to define the quantum of thermal conductance as
$\kappa_0 = L_0 T G_0$. If we use $\kappa_0$ in the following, we will set the
absolute temperature to $T=315$~K, consistently with our choice in
Sec.~\ref{subsec:electron-transport}.

Subsequently we will use the electronic Lorenz ratio,
\begin{equation} \label{eqn:lorenzratio-el}
  \frac{L_{\text{el}}}{L_0} = \frac{\kappa_{\text{el}}}{L_0 T G},
\end{equation}
to discuss the validity of the Wiedemann-Franz law for our simulations. A
value of $L_{\text{el}}/L_{0} = 1$ indicates perfect agreement with it and
considers only electronic transport quantities $G$ and
$\kappa_{\text{el}}$. Since in the experiments only the total thermal
conductance $\kappa=\kappa_{\text{el}}+\kappa_{\text{ph}}$ can be measured, we
also introduce the Lorenz ratio,
\begin{equation} \label{eqn:lorenzratio}
  \frac{L}{L_0} = \frac{\kappa}{L_0 T G}
\end{equation}
to quantify deviations from a ``generalized'' Wiedemann-Franz law $\kappa
= L_{0} T G$ due to both electrons and phonons.

\section{\label{sec:results}Results and discussion}

In this section we present the simulation results for the three materials Au,
Pt, and Al.  We performed 60 stretching events for atomic junctions of each
metal. Different junction evolutions were obtained by selecting random initial
velocities of wire and thermal reservoir atoms. After discussing exemplarily a
single stretching process, we will analyze the 60 junction evolutions and
related transport properties statistically.

\subsection{\label{subsc:gold}Gold}

\subsubsection{Single stretching process}

\begin{figure}
  \includegraphics[width=0.5\textwidth]{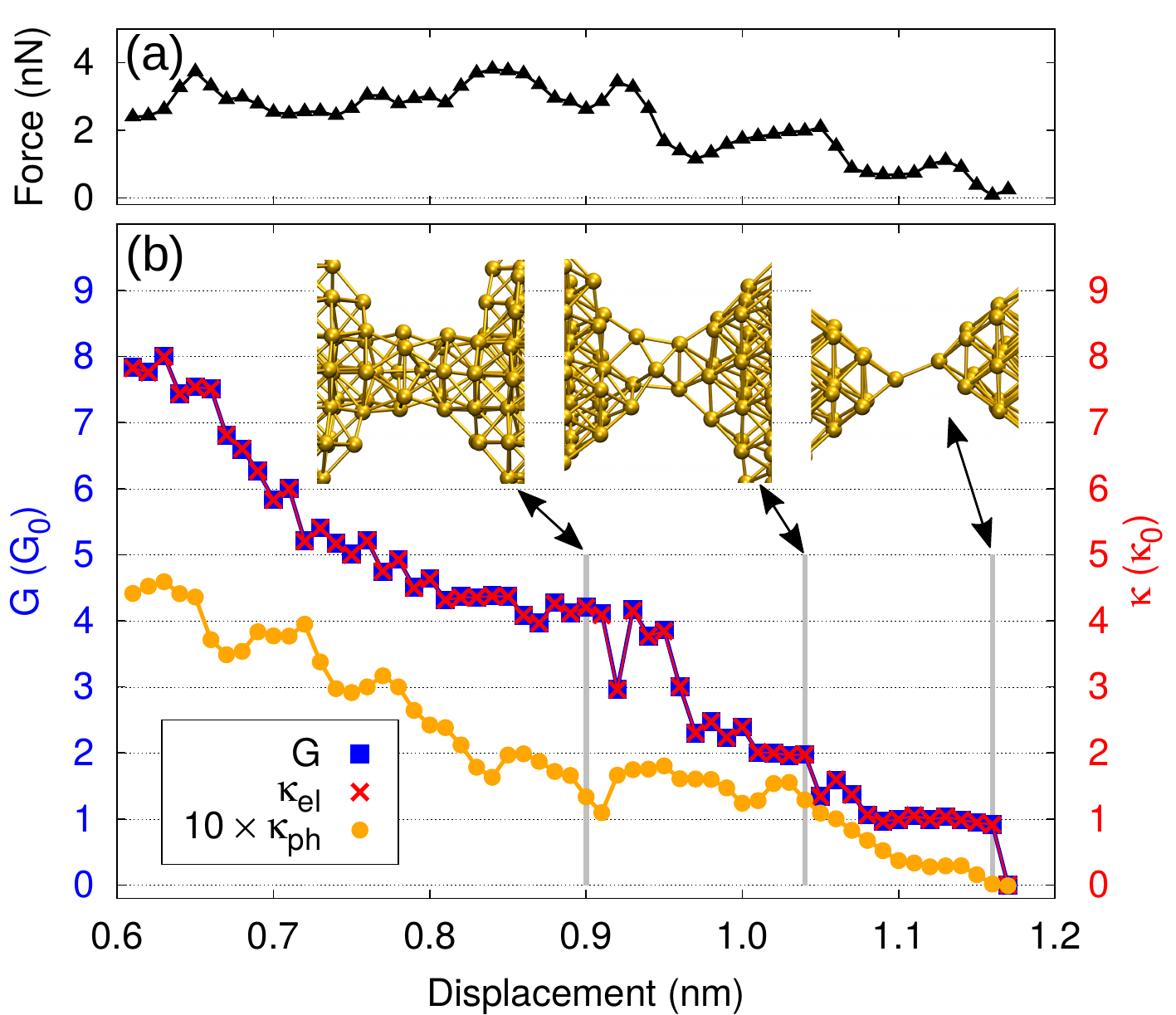}
  \caption{Single stretching process for Au. (a) Force acting on the fixed
    atoms of the right side (see Fig. \ref{fig:md_geometry}) as a function of
    electrode displacement. (b) Electrical conductance $G$ (blue squares) and
    electronic as well as phononic thermal conductance $\kappa_{\text{el}}$
    (red crosses) and $\kappa_{\text{ph}}$ (orange circles) as a function of
    electrode displacement. Note that
    the phononic thermal conductance is magnified by a factor of 10. The three
    insets show the narrowest part of the wire at displacements of $0.90$~nm,
    $1.04$~nm, and $1.16$~nm, as indicated by vertical gray lines.}
  \label{fig:single_gold}
\end{figure}

Figure~\ref{fig:single_gold} shows a single stretching process for a gold
nanojunction. Figure~\ref{fig:single_gold}(a) presents the sum of all forces
acting in the direction of the stretching on the fixed atoms that surround the
cold reservoir. Plastic and elastic stages can be distinguished. During the
elastic stages forces build up, which decrease in plastic stages, when atomic
bonds break and atoms in the junction rearrange. The conductance $G$ (blue
squares), electronic thermal conductance $\kappa_{\text{el}}$ (red crosses),
and the tenfold magnified phononic thermal conductance $\kappa_{\text{ph}}$
(orange dots) versus electrode displacement are depicted in
Fig.~\ref{fig:single_gold}(b). The electrical conductance exhibits a trace
with typical features seen in experimental results \cite{costa1997,
  muller1996}.  Before junction rupture at a displacement of $1.16$~nm, a
plateau at $1G_0$ can be observed.  This plateau is a result of the atomic
configuration shown as an inset on the right. This dimer configuration
consists basically of a two-atom chain formed from the tip atoms of two
atomically sharp electrode pyramids that touch each other. Before this last
plateau, a shorter plateau at roughly $2G_{0}$ is found. The middle inset
shows the corresponding atomic configuration.  At an even smaller electrode
separation, we see a plateau at a conductance of $4G_0$. The left inset
relates this to a structure with two pentagonal rings and a single atom
between them.

Measuring $G$ and $\kappa_{\text{el}}$ in terms of their respective
conductance quanta, we find a nearly perfect match between both quantities
throughout the whole stretching process. This is expected from the
Wiedemann-Franz law in Eq.~(\ref{eqn:WFlaw}), if the transmission function
$\tau(E)$ is weakly energy-dependent around the Fermi energy, as it is
typically the case for Au atomic contacts \cite{kloeckner2017,buerkle2018}.

The phononic thermal conductance $\kappa_{\text{ph}}$ decreases overall within
the simulation run. For structures with large narrowest cross sections at
  small displacements at the beginning of the simulations (see the plot and
  discussion in Sec.~\ref{sec:FullPull-ComputationalCost}), it shows a clear
  saw-tooth-like profile. During elastic displacements of atoms, the force
increases steadily, while the phononic thermal conductance tends to
decrease. For a displacement large enough, the structure then reconfigures,
reducing the force and abruptly increasing the phononic thermal conductance.
This can be understood by considering the force constants, which are changing
with interatomic distance for anharmonic potentials. The decreasing force
constants lead to a decrease in the phononic thermal conductance until they
are restored again after the reconfiguration \cite{kloeckner2017}. For the
rather elongated structures shown in Fig.~\ref{fig:single_gold} this behavior
is, however, not very pronounced.

Between a displacement of $1.0$~nm and the breaking of the wire at 1.16~nm,
two steps in the force trace can be observed.  The first step is accompanied
by a reconfiguration of the structure with a conductance of $2G_0$ to the
dimer structure with a conductance of $1G_0$. $\kappa_{\text{ph}}$ follows
this decline in the force more smoothly than $\kappa_{\text{el}}$. The second
step in the force is due to the breaking of the wire.  While the electronic
transport properties drop to zero from a plateau at $1G_{0}$ or $1
\kappa_{0}$, respectively, $\kappa_{\text{ph}}$ decreases rather smoothly. We
note that the behavior of $\kappa_{\text{ph}}$ in this case resembles closely
the force, which shows a rather continuous decrease before the point of
rupture in contrast to the sharp drop of the electronic transport
properties. This can be understood by recalling that in the case of a dimer
contact the force and $\kappa_{\text{ph}}$ will basically be determined by the
mechanical interaction between the two tip atoms.

The study of Fig.~\ref{fig:single_gold} reveals that $\kappa_{\text{ph}}$
reacts somewhat differently to certain atomic reconfigurations than
$\kappa_{\text{el}}$ (or, equivalently, $G$), as discussed in the previous
paragraph. Another example can be seen in the range of displacements between
$0.80$ and $1.04$~nm. Here the electronic thermal conductance
$\kappa_{\text{el}}$ stays roughly constant at a value of around
$4.5\kappa_{0}$ and then drops rapidly towards $2\kappa_{0}$, with the
exception of the dip at 0.92~nm. In contrast $\kappa_{\text{ph}}$ shows two
smaller dips but then remains roughly constant during the rapid drop of
$\kappa_{\text{el}}$. Fundamentally this behavior originates from differences
in the sensitivity of mechanical interatomic interactions and electronic
structure to atomic junction configurations.

\subsubsection{Statistical analysis}

\begin{figure}
  \includegraphics[width=0.5\textwidth]{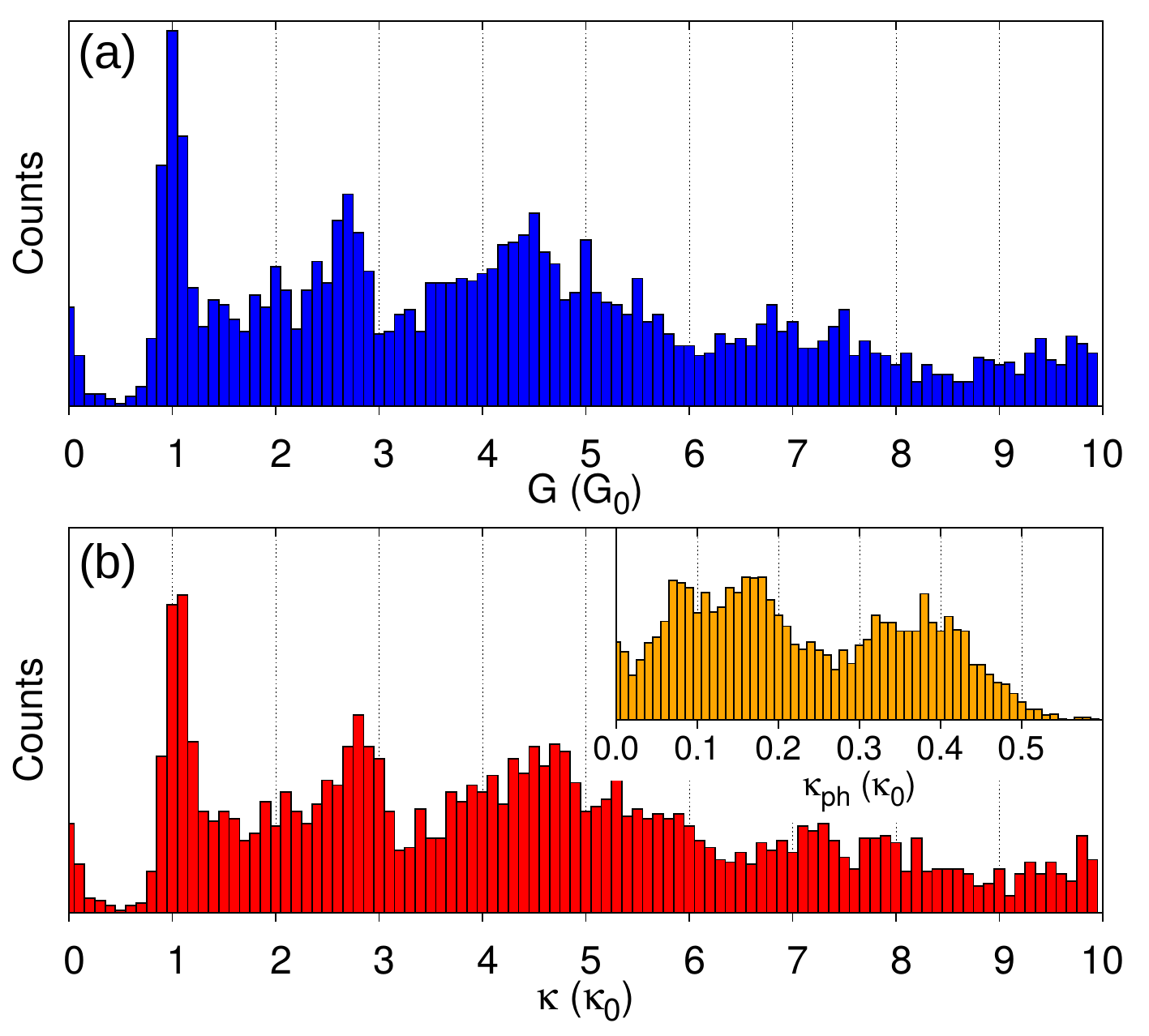}
  \caption{Histograms for Au for (a) the electrical conductance $G$, (b) the
    total thermal conductance $\kappa = \kappa_{\text{el}} +
    \kappa_{\text{ph}}$, and in the inset for $\kappa_{\text{ph}}$. We use bin
    sizes of $0.1G_0$ and $0.1\kappa_0$ for electrical and thermal
    conductances, respectively.}
  \label{fig:statistic_gold_1d}
\end{figure}

\begin{figure*}
  \includegraphics[width=\textwidth]{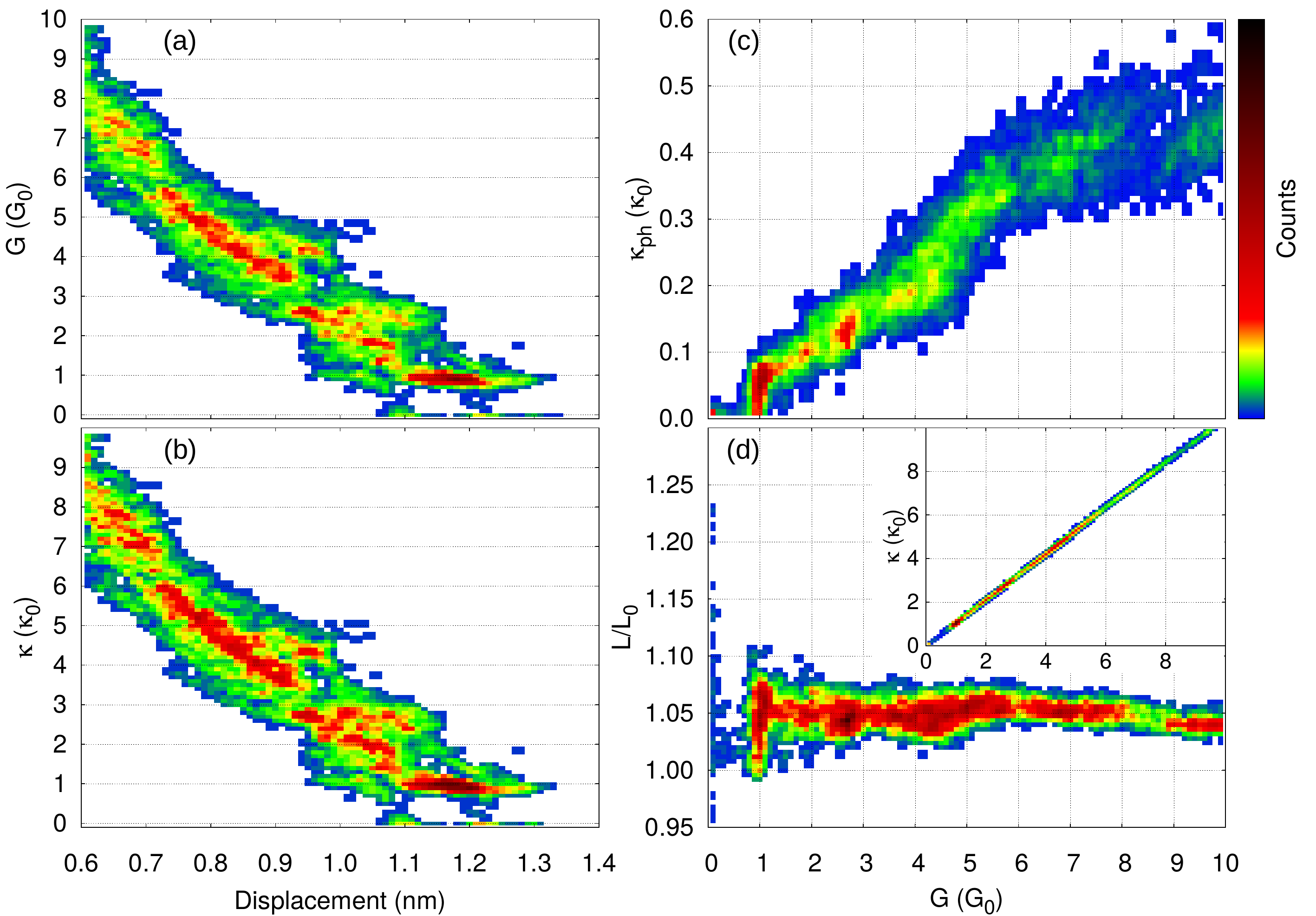}
  \caption{Density plots for Au. (a) Electrical conductance $G$ and (b)
    thermal conductance $\kappa$ over electrode displacement,
    respectively. (c) Phononic thermal conductance $\kappa_{\text{ph}}$ and
    (d) Lorenz ratio $L/L_{0}$ as well as in the inset thermal conductance
    $\kappa$ over electrical conductance $G$, respectively.  For the density
    plots we use bin sizes of $0.1G_0$ and $0.1\kappa_{0}$ for the electrical
    and thermal conductances, $0.004$ for the Lorenz ratio, and 0.01~nm for
    the displacement bins, corresponding to 1~ns at the chosen electrode
    separation speed.}
  \label{fig:statistic_gold_2d}
\end{figure*}

Let us now take a look at the results of $60$ stretching processes, each of
them similar to the simulation run described in the previous paragraph. We
will analyze these results statistically in the following.

Figure~\ref{fig:statistic_gold_1d}(a) shows the histogram of the electrical
conductance $G$, Fig.~\ref{fig:statistic_gold_1d}(b) those of the total
thermal conductance $\kappa$ and in the inset those of the phononic thermal
conductance $\kappa_{\text{ph}}$. The histogram for $G$ exhibits a large peak
at $1G_0$, which corresponds to the plateau at $1G_0$ for the single
conductance-distance trace in Fig.~\ref{fig:single_gold}(b) and is in
agreement with experimental results \cite{costa1997, muller1996, cui2017}. A
small peak at $2G_0$ is hardly discernible. Instead it is more likely in our
simulations to find configurations with conductances centered around
$2.7G_0$. Another area of increased counts is visible at $4.5G_0$. It stems
mainly from the structures with two pentagonal rings interconnected by a
single atom, discussed in Fig.~\ref{fig:single_gold}.

The thermal conductance histogram for $\kappa$ closely resembles those of the
electrical conductance $G$. This can be anticipated from the single stretching
trace, which manifests near perfect proportionality between $G$ and
$\kappa_{\text{el}}$, as expected from the Wiedemann-Franz law in
Eq.~(\ref{eqn:WFlaw}), and only a minor contribution of
$\kappa_{\text{ph}}$. Figure~\ref{fig:single_gold} indeed shows that
$\kappa_{\text{ph}}$ is more than an order of magnitude smaller than
$\kappa_{\text{el}}$. In detail we find that peaks in the thermal conductance
histogram are shifted slightly towards larger values than for $G$, which we
attribute basically to the phononic contribution.

The phononic thermal conductance $\kappa_{\text{ph}}$ in the inset of
Fig.~\ref{fig:statistic_gold_1d}(b) shows no clear peaks.  Instead there are
two broad intervals of high counts centered around $0.13\kappa_0$ and
$0.37\kappa_0$. The latter region results from junctions at the beginning of
the pulling process shown in Fig.~\ref{fig:single_gold}. The area around
$0.13\kappa_0$, on the other hand, originates from a nearly constant
$\kappa_{\text{ph}}$ while $G$ changes from more than $4G_0$ to less than
$2G_0$, seen again in Fig.~\ref{fig:single_gold}. The counts at values smaller
than $0.05\kappa_0$ stem mainly from the last electrical conductance plateau
at $1G_0$ and the smooth decrease of $\kappa_{\text{ph}}$ shortly before the
breaking of the atomic wire.

In Fig.~\ref{fig:statistic_gold_2d} density plots of the different transport
quantities are shown. Figure~\ref{fig:statistic_gold_2d}(a) displays the density
plot for conductance $G$ over displacement. We find mainly three areas of
increased counts centered at $1G_0$, $2.7G_0$ and $4.5G_0$, as expected from
the one-dimensional histogram in Fig.~\ref{fig:statistic_gold_1d}(a). The
  corresponding preferred configurations for gold are bridges consisting of
pentagonal rings connected by a single atom [left inset in
  Fig.~\ref{fig:single_gold}(b)], structures with only two or three atoms in
the narrowest part [middle inset in Fig.~\ref{fig:single_gold}(b)], and
single-atom contacts [right inset in Fig.~\ref{fig:single_gold}(b)].  The
transitions between these configurations happen relatively fast on a nanosecond time
scale.

Figure~\ref{fig:statistic_gold_2d}(b) shows the total thermal conductance
$\kappa$ over the displacement.  As expected from the histograms in
Fig.~\ref{fig:statistic_gold_1d}, the differences between $G$ and $\kappa$ are
only minor and mainly stem from the small phononic contribution
$\kappa_{\text{ph}}$. Since all the major features of
Fig.~\ref{fig:statistic_gold_2d}(a) can be recovered, we refrain from
discussing them again.

Figure~\ref{fig:statistic_gold_2d}(c) plots $\kappa_{\text{ph}}$ over $G$. While
the electrode displacement is a measure for the simulation time, $G$ is an
approximate measure of the narrowest cross section of the contact. We find
that $\kappa_{\text{ph}}$ shows a weak decrease with decreasing $G$ for large
values ($G>6G_0$).  The decrease becomes larger for $4G_0<G<6G_0$ in the range
of the pentagonal ring structures. The increased slope suggests a high
rigidity of these bridges, since $\kappa_{\text{ph}}$ decreases faster with
$G$ than for other structures.  For smaller $G$, the decrease of
$\kappa_{\text{ph}}$ flattens again until $\kappa_{\text{ph}}$ drops to zero
rapidly at $1G_0$.

In Fig.~\ref{fig:statistic_gold_2d}(d) the density plot of the Lorenz ratio
$L/L_{0}$ is depicted. For the largest structures at $10G_0$ it is centered
around $1.04$ and increases slightly to $1.05$ for smaller structures between
$1G_0$ and $6G_0$. The variation of $L/L_0$ increases for narrower wires, and,
although still moderate with $1\leq L/L_0\leq 1.1$, in the contact regime it
is largest at $1G_0$. As can be seen in Fig.~\ref{fig:single_gold}, while $G$
remains in the plateau at $1G_0$, $\kappa_{\text{ph}}$ varies from
$0.1\kappa_{0}$ to $0$, which nicely explains the range of variation of
$L/L_0$. We have attained only few counts in the tunneling regime
(corresponding roughly to the conductance regime $0 \lesssim G/G_0 \lesssim
0.5$), so that these phononic properties cannot be discussed reliably. It
is a regime of strong current interest
\cite{kim2015,cui2017_2,kloppstech2017}, and further theoretical studies are
desirable that take into account also radiative contributions to the heat
transport. Finally, the inset of Fig.~\ref{fig:statistic_gold_2d}(d) shows the
total thermal conductance $\kappa$ over the conductance $G$.  In accordance
with the results for the Lorenz ratio, $\kappa$ increases linearly with $G$ at
a slope of $L/L_0\cdot\kappa_0/G_0$.

To summarize, since the electronic transport quantities $G$ and
$\kappa_{\text{el}}$ show excellent agreement with the Wiedemann-Franz law,
the average deviations of $L/L_0$ from 1 on the order of $5\%$ can be
attributed mainly to phononic contributions. This is compatible with our
conclusions in Refs.~\cite{cui2017,kloeckner2017} based on a few selected
geometries described by density functional theory.

\subsection{\label{subsc:platinum}Platinum}

\subsubsection{Single stretching process}

\begin{figure}[t!]
  \includegraphics[width=0.5\textwidth]{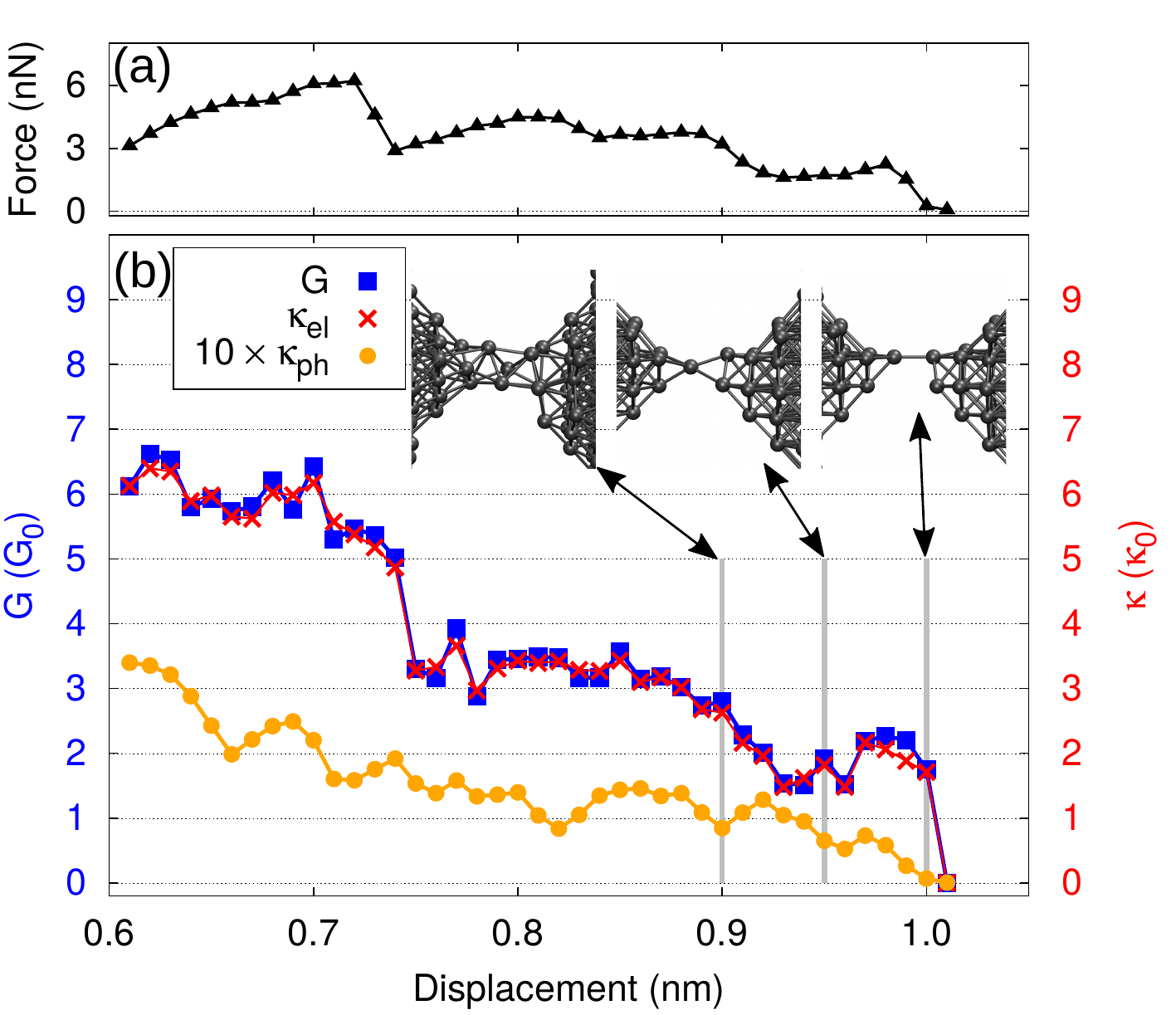}
  \caption{Same results as Fig.~\ref{fig:single_gold} but for Pt. The three
    insets show the narrowest part of the wire at displacements of $0.90$~nm,
    $0.95$~nm and $1.00$~nm, as indicated by vertical gray lines.}
  \label{fig:single_platinum}
\end{figure}

In analogy to Fig.~\ref{fig:single_gold}, Fig.~\ref{fig:single_platinum} shows
a single junction stretching process for platinum. The force in
Fig.~\ref{fig:single_platinum}(a) is higher than for similar structures of
gold, as expected, since platinum is more rigid and has a higher melting point
\cite{kaye1973, ashcroft1976}.  Figure~\ref{fig:single_platinum}(b) shows the
electronic and phononic transport properties $G$, $\kappa_{\text{el}}$ and
$\kappa_{\text{ph}}$ as a function of electrode displacement. We observe that
$G$ and $\kappa_{\text{el}}$ agree less well than for gold, signaling purely
electronic deviations from the Wiedemann-Franz law.  This was also observed in
Refs.~\cite{cui2017,kloeckner2017} and can be attributed to the stronger
energy dependence of the transmission at the Fermi energy due to $d$
states. Similar to Au, $\kappa_{\text{ph}}$ shows a somewhat smoother behavior
than $\kappa_{\text{el}}$. For instance the drop in $\kappa_{\text{el}}$ from
$5\kappa_0$ to $3.2\kappa_0$ is not matched by $\kappa_{\text{ph}}$. Also the
steady decrease of $\kappa_{\text{el}}$ around a displacement of $0.9$~nm is
not observed in $\kappa_{\text{ph}}$, which just shows a small dip but an
otherwise rather constant value. Since the $d$ orbitals of platinum, which are
not spherically symmetric, contribute to the electronic transport
\cite{muller1992, pauly2006}, the sensitivity of electronic transport
properties to geometric changes is enhanced as compared to the phononic
properties, which are derived from spherically symmetric EAM potentials. The
last plateau before breaking exhibits a conductance of approximately $2G_0$ as
in Ref.~\cite{muller1992}. It is higher than for gold, which is again assigned
to the $d$ orbitals that contribute to electronic transport in addition to the
$s$ orbitals. Similarly to gold, we observe that $\kappa_{\text{ph}}$ vanishes
in a rather smooth fashion that already sets in before $\kappa_{\text{el}}$
drops to zero abruptly.

\subsubsection{Statistical analysis}

\begin{figure}[t!]
  \includegraphics[width=0.5\textwidth]{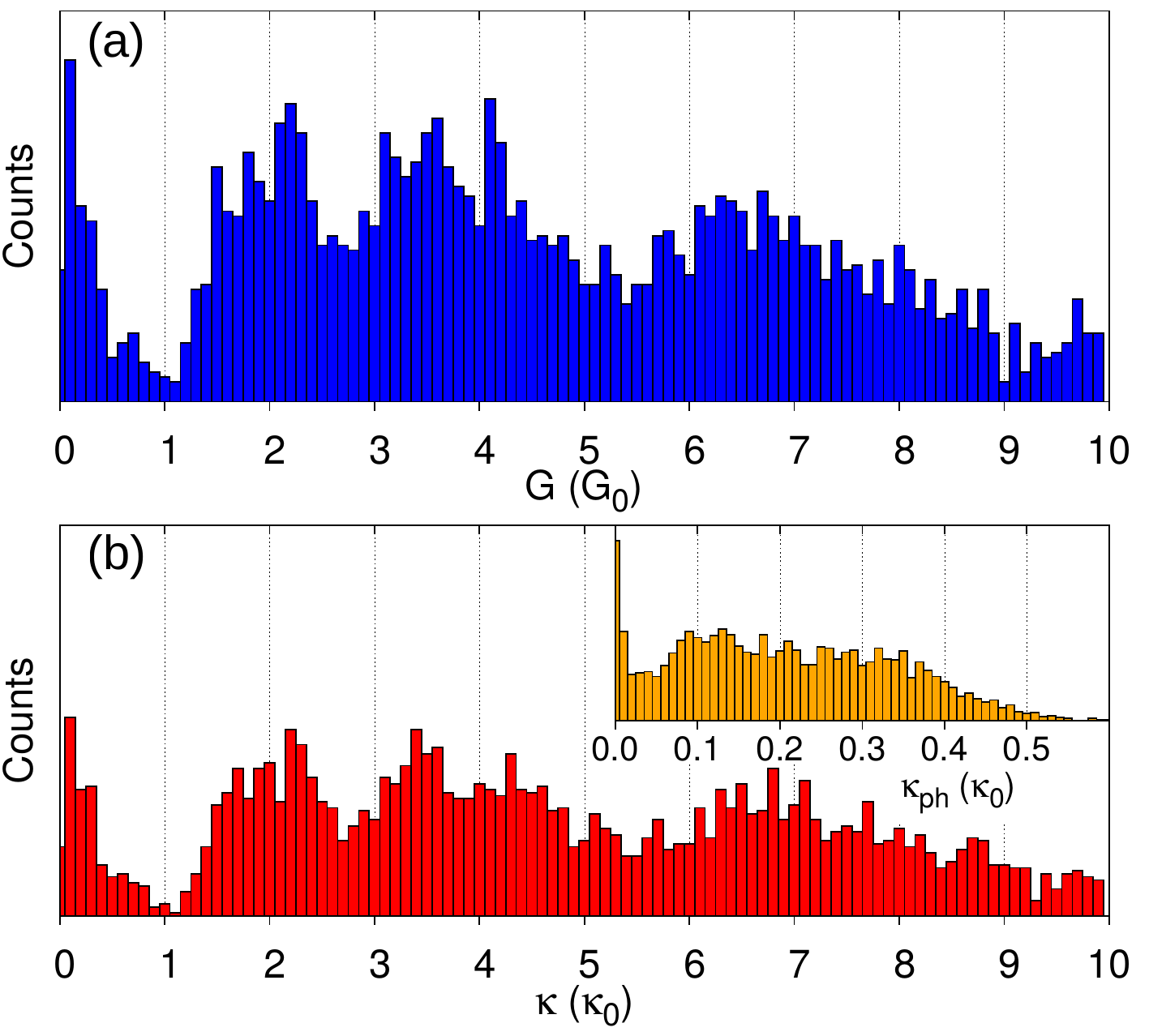}
  \caption{Histograms as in Fig.~\ref{fig:statistic_gold_1d} but for Pt.}
  \label{fig:statistic_platinum_1d}
\end{figure}

\begin{figure*}[t!]
  \includegraphics[width=\textwidth]{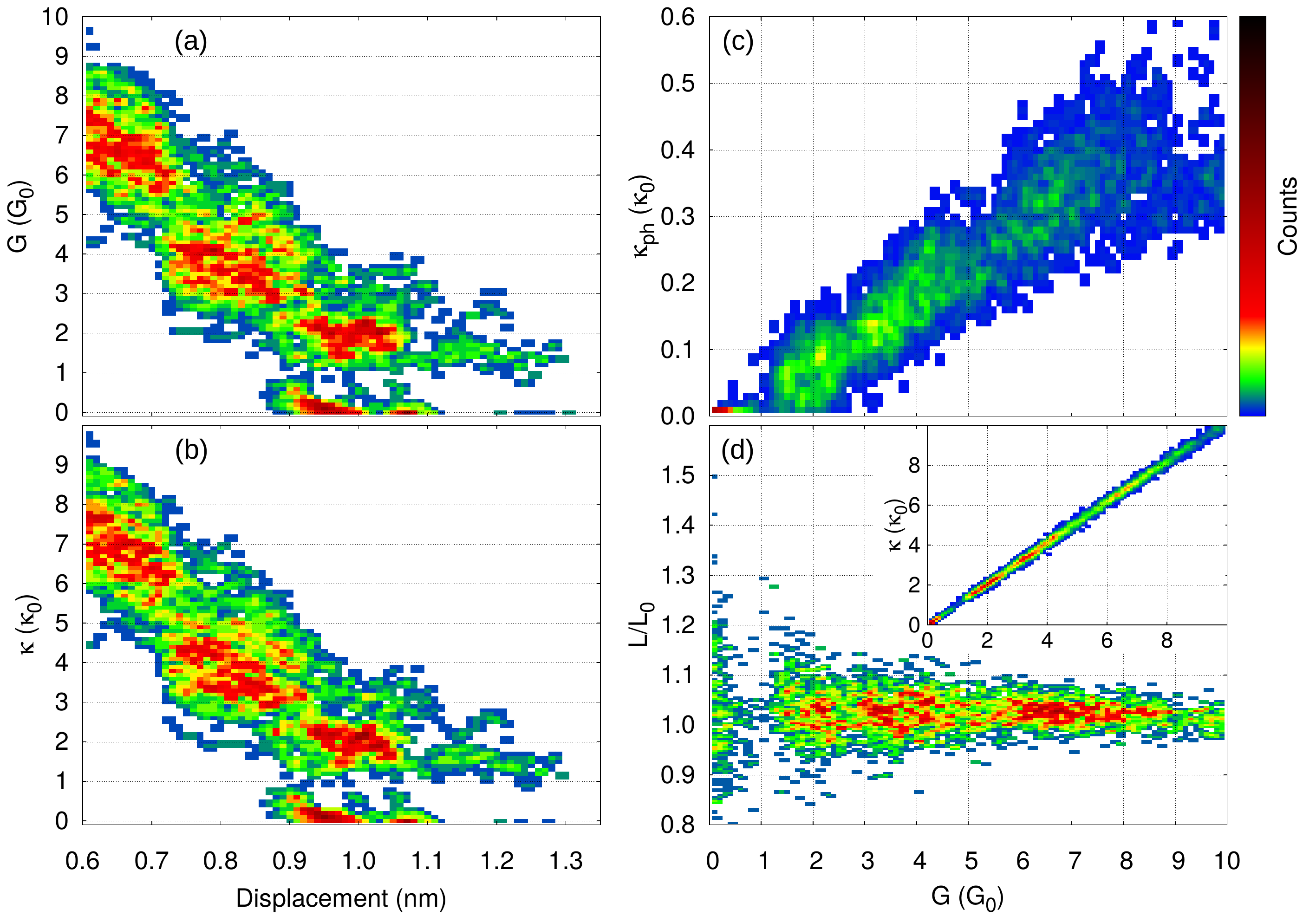}
  \caption{Density plots as in Fig.~\ref{fig:statistic_gold_2d} but for Pt.}
  \label{fig:statistic_platinum_2d}
\end{figure*}

Now we take a look at the statistical results for
platinum. Figure~\ref{fig:statistic_platinum_1d} shows the conventional
conductance histograms for $G$, $\kappa$ and $\kappa_{\text{ph}}$.  The
histogram for $G$ in Fig.~\ref{fig:statistic_platinum_1d}(a) features regions
of increased counts around $2G_0$, $3.6G_0$ and $6.5G_0$. As can be inferred
from the single trace in Fig.~\ref{fig:single_platinum}, the region at around
$2G_0$ corresponds to single-atom contacts in monomer (middle inset) and dimer
configuration (right inset). The second region, centered at $3.6G_0$, relates
to configurations with two atoms in the narrowest part, as shown in the left
inset of Fig.~\ref{fig:single_platinum}, while structures at around $6.5G_0$
exhibit typically four atoms at the narrowest cross section. As for Au, the
histogram of the total thermal conductance $\kappa$ in
Fig.~\ref{fig:statistic_platinum_1d}(b) shows a similar shape as those of
$G$. The histogram of the phononic thermal conductance $\kappa_{\text{ph}}$ is
rather featureless.

In Fig.~\ref{fig:statistic_platinum_2d} the density plots of the various
transport quantities are displayed. In Figs.~\ref{fig:statistic_platinum_2d}(a)
and ~\ref{fig:statistic_platinum_2d}(b) the conductance $G$ and the total
thermal conductance $\kappa$ over the displacement show only small
differences.  In comparison to gold a larger spread in the transport
properties at a given displacement is evident for platinum.

Figure~\ref{fig:statistic_platinum_2d}(c) shows the two-dimensional histogram of
phononic thermal conductance $\kappa_{\text{ph}}$ versus electrical
conductance $G$. In contrast to the results for gold, $\kappa_{\text{ph}}$
increases rather linearly with $G$ for all conductances, although with a wide
spread.

This spread is also evident from the Lorenz ratio in
Fig.~\ref{fig:statistic_platinum_2d}(d). Being centered around 1.05 due to a
5\% offset from 1 by phononic effects, it can vary from $0.9\lesssim
L/L_0\lesssim 1.2$ for the single-atom contact configurations. Opposite to Au
electronic effects may contribute up to 10\% to variations around the mean in
Pt, as can be seen from the electronic Lorenz ratio $L_{\text{el}}/L_0$
\cite{kloeckner2017}. Another 10\% of variations comes from the phonons,
adding up to a maximum of 20\% difference from the Wiedemann-Franz law. The
inset in Fig.~\ref{fig:statistic_platinum_2d}(d) visualizes the linear
relationship between $\kappa$ and $G$.

\subsection{\label{subsc:aluminium}Aluminum}

\subsubsection{Single stretching process}

\begin{figure}
  \includegraphics[width=0.5\textwidth]{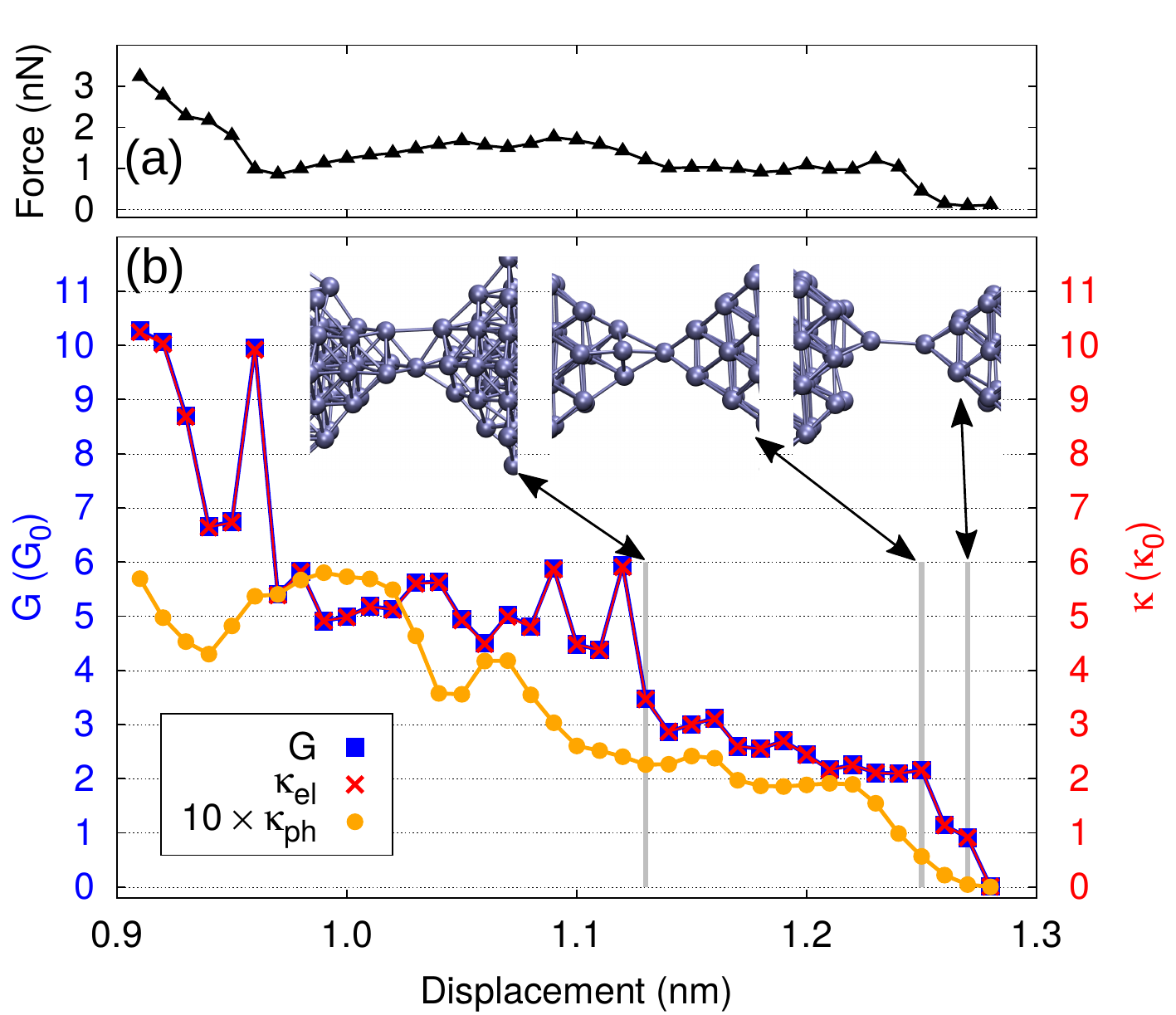}
  \caption{Same results as in Fig. \ref{fig:single_gold} but for Al. The three
    insets show the narrowest part of the wire at displacements of $1.13$~nm,
    $1.25$~nm and $1.27$~nm, as indicated by vertical gray lines.}
  \label{fig:single_aluminium}
\end{figure}

In Fig.~\ref{fig:single_aluminium} the results of a single stretching process
for an aluminum junction are plotted. Figure~\ref{fig:single_aluminium}(a) shows
the force to be smaller than for gold.  In Fig.~\ref{fig:single_aluminium}(b),
the conductance $G$ and the electronic thermal conductance
$\kappa_{\text{el}}$ manifest a high degree of correlation, meaning that the
Wiedemann-Franz law is well fulfilled. Both $G$ and $\kappa_{\text{el}}$ show
large fluctuations. We attribute them to the relative softness of the material
compared to gold and platinum. This can be quantified by the lowest Young's
modulus in the set of metals studied or the lowest melting point
\cite{kaye1973, ashcroft1976}.  For this reason atomic reconfigurations occur
easily at relatively small forces. As compared to Au, $p$ orbitals contribute
strongly to charge transport beside $s$ orbitals. Similar to the $d$ orbitals
in platinum, the $p$ orbitals are not spherically symmetric and hence the
electronic structure is quite sensitive to changes in the atomic
configurations.

As previously observed for Au and Pt, the phononic thermal conductance
$\kappa_{\text{ph}}$ shows a relatively smooth decay and no clear correlation
with $\kappa_{\text{el}}$. For instance, while $\kappa_{\text{el}}$ decreases
significantly from $10\kappa_0$ to below $6\kappa_0$ in a range of
displacements from 0.9 to 1~nm, $\kappa_{\text{ph}}$ decreases to a local
minimum and rises again to a similar value as before. And around the
displacement of $1.1$~nm $\kappa_{\text{el}}$ shows two peaks, while
$\kappa_{\text{ph}}$ decreases smoothly. We observe again that
$\kappa_{\text{ph}}$ drops to zero faster than $\kappa_{\text{el}}$, as is
evident before the rupture of the contact.

The middle inset in Fig.~\ref{fig:single_aluminium}(b) shows a monomer
structure, which yields a conductance of $2G_0$. It is realized throughout the
long plateau at $2G_0$, as also suggested by the leftmost geometry in the
inset of Fig.~\ref{fig:single_aluminium}(b) at the beginning of the
plateau. We find only two data points with a conductance of about $1G_0$ for
the dimer structure shown as the right inset at a displacement of 1.27~nm.

\subsubsection{Statistical analysis}

\begin{figure}[t!]
  \includegraphics[width=0.5\textwidth]{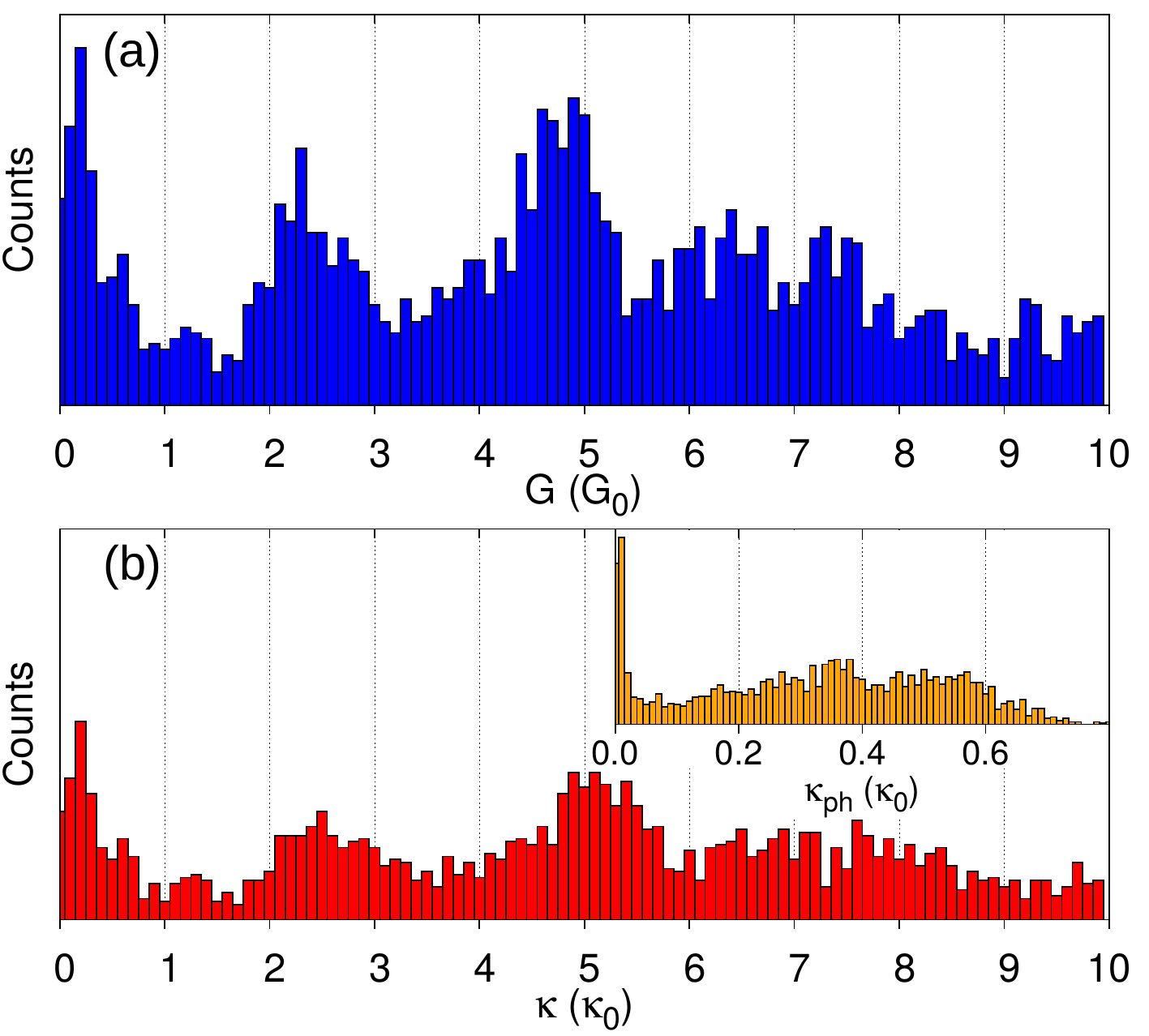}
  \caption{Histograms as in Fig.~\ref{fig:statistic_gold_1d} but for Al.}
  \label{fig:statistic_aluminium_1d}
\end{figure}

\begin{figure*}[t!]
  \includegraphics[width=\textwidth]{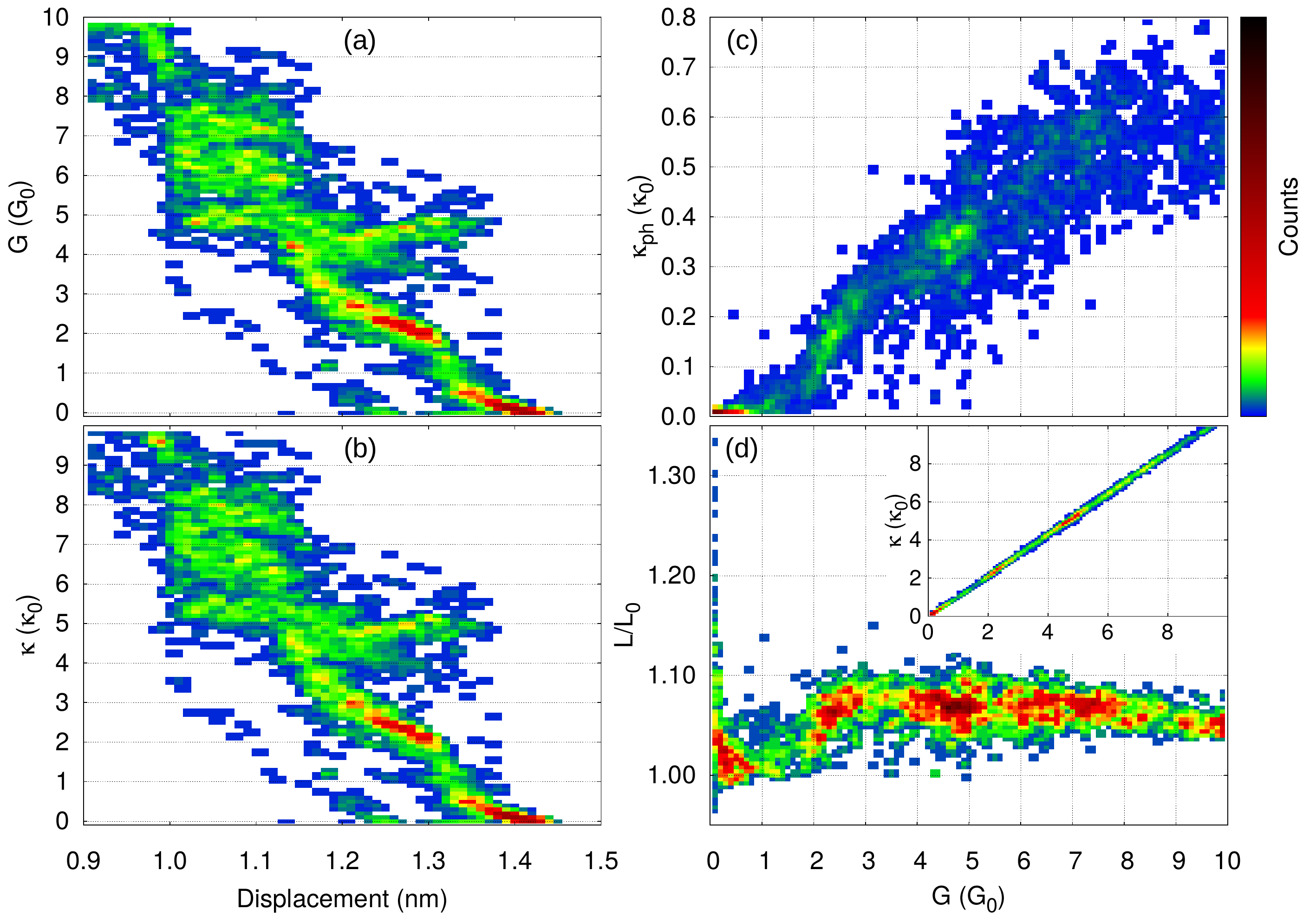}
  \caption{Density plots as in Fig.~\ref{fig:statistic_gold_2d} but for Al.}
  \label{fig:statistic_aluminium_2d}
\end{figure*}

Figure~\ref{fig:statistic_aluminium_1d} depicts the histograms for the
conductance $G$, the total thermal conductance $\kappa$, and the phononic heat
conductance $\kappa_{\text{ph}}$.  In the histogram for $G$ in
Fig.~\ref{fig:statistic_aluminium_1d}(a) the counts centered at $1.2G_0$ can
be assigned to dimer configurations, as shown in the right inset of
Fig.~\ref{fig:single_aluminium}(b). Due to the relative instability of these
dimer configurations at room temperature, the number of data points at this
conductance is very low. Let us note that the absence of a peak in the
electrical conductance histogram around $1G_0$ is at variance with
experimental findings at low temperature \cite{yanson1997}. The instability of
the dimer structures in our room-temperature simulations may be a shortcoming
of the chosen standard EAM potential. A larger peak at $2.3G_0$ originates
from monomer configurations, similar to that shown in the middle inset in
Fig.~\ref{fig:single_aluminium}(b). The increased count number at $4.8G_0$
stems from atomic configurations with pentagonal rings and a single atom in
the middle, as discussed for Au before.

The histogram for $\kappa$ in Fig.~\ref{fig:statistic_aluminium_1d}(b)
possesses the same overall shape as the histogram for $G$. The histogram for
the phononic thermal conductance $\kappa_{\text{ph}}$ in the inset of
Fig.~\ref{fig:statistic_aluminium_1d}(b) shows no distinct features but a
broad distribution. The maximum of counts around $0.36\kappa_{0}$ is related
to the pentagonal ring configurations.

The analysis in terms of density plots is depicted in
Fig.~\ref{fig:statistic_aluminium_2d}. The conductance $G$, which is related
to the displacement in Fig.~\ref{fig:statistic_aluminium_2d}(a), shows a large
variation for the different simulation runs in terms of the time point, at
which contacts narrow and finally break. A large area of increased counts
between displacements of $1.0$ and $1.4$~nm at a conductance of around
$4.8G_0$ illustrates the high probability of formation of the stable
pentagonal ring configurations. It is also well visible that contacts tend to
break at a large cross section and before the dimer structures with a
conductance of around $1G_0$ are realized.

The total thermal conductance $\kappa$ in
Fig.~\ref{fig:statistic_aluminium_2d}(b) shows a very similar behavior as $G$
due to a small $\kappa_{\text{ph}}$ and the Wiedemann-Franz law. We therefore do
not discuss it any further.

Figure~\ref{fig:statistic_aluminium_2d}(c) relates $\kappa_{\text{ph}}$ and $G$.
An interesting feature is that $\kappa_{\text{ph}}$ stays basically below
$0.1\kappa_0$ for $G$ between $1G_0$ and $2G_0$. At roughly $2G_0$ the
phononic thermal conductance suddenly increases rapidly to nearly
$0.2\kappa_0$.  This suggests a high sensitivity of $\kappa_{\text{ph}}$ to
reconfigurations of single-atom contacts but we can also relate this
observation to the instability of dimer structures, for which the phononic
transport is clearly reduced.  For $G>2.5G_0$, $\kappa_{\text{ph}}$ increases
linearly.

The Lorenz ratio in Fig.~\ref{fig:statistic_aluminium_2d}(d) is centered
between $1.00$ and $1.05$ for $G < 2G_0$ and between $1.05$ and $1.08$ for
$G\geq 2G_{0}$. In Ref.~\cite{kloeckner2017} we have pointed out that the
phononic contribution $\kappa_{\text{ph}}$ may reach 40\% of
$\kappa_{\text{el}}$ for a dimer configuration. A main reason for this
prediction is a possible suppression of the electronic transport in such
junction geometries, leading to $G\approx0.5G_0$ and a correspondingly small
$\kappa_{\text{el}}$. We could not reproduce this finding here, as the dimer
junctions were too unstable to yield a statistically significant amount of
data points. The inset of Fig.~\ref{fig:statistic_aluminium_2d}(d)
demonstrates the good linear relationship between $\kappa$ and $G$ with a
slope proportional to $L/L_0$.

\section{\label{sec:conclusions}Conclusions}

In this paper we have introduced a new computational scheme to study the
phonon thermal conductance of nanojunctions. It is general and can be applied
to any nanosystem, for which interatomic potentials are known. By combining
the NEMD technique with a tight-binding parametrization and the
Landauer-B\"uttiker formalism, we have illustrated its use for the description
of transport properties of metallic atomic contacts by investigating the
electrical conductance as well as electronic and phononic contributions to the
thermal conductance. Motivated by recent experimental tests of the
Wiedemann-Franz law at the atomic scale and at room temperature
\cite{cui2017,mosso2017,mosso2019}, our main focus was to estimate the
contributions of phonons and electrons to the heat conductance. Our method
proved to be computationally fast enough to perform several ten to hundred
stretching simulations for atomic contacts consisting of hundreds of
atoms. Detailed requirements of computational resources for the presented
  study are reported in Sec.~\ref{sec:FullPull-ComputationalCost}.

Concerning the validity of the Wiedemann-Franz law in metallic atomic
contacts, for Au and Pt we showed that at room temperature mean deviations due
to phonons are around 5\%, in agreement with a previous
density-functional-theory-based analysis \cite{kloeckner2017} and with
experimental findings \cite{cui2017}. Maximum deviations in the contact region
were below 10\% for Au mainly due to phonons but could range up to 20\% for
Pt. Electronic and phononic effects are responsible for this enhanced
deviation in Pt to a similar extent. For Al atomic junctions we are not aware
of any measurements of the thermal conductance, and our calculations should
thus be seen as predictions. In this case, we found deviations from the
Wiedemann-Franz law to be on average below 8\% with maximum deviations up to
around 10\%. Previous predictions of substantial deviations of up to 40\% by
some of the present authors \cite{kloeckner2017} could not be confirmed due to
an instability of dimer junction geometries at room temperature with the
chosen EAM potential.

We observe a difference in the sensitivity to atomic reconfigurations between
electronic and phononic thermal transport. Especially for Pt and Al contacts,
the electronic thermal conductance shows more variation than the phononic
thermal conductance. We attribute this to additional contributions of $d$ and
$p$ orbitals to transport, respectively, which are not spherically symmetric
in contrast to the EAM potentials used. We also find a rapid decrease in the
phononic thermal conductance contributions before the final breaking of atomic
wires, as signaled by a sharp drop in the electrical conductance. We attribute
it to a weakening of the mechanical coupling within the wires that sets in
already before electronic overlap is lost and indicates that an atomic contact
becomes unstable.

In conclusion, the computational approach presented in this paper reproduces
experimental findings for different metals and can be used to make predictions
for unexplored materials. At the same time it is computationally efficient
enough to facilitate a statistical analysis of charge and heat transport
properties.

As an outlook, to further improve the computational results, a careful
  assessment of the performance of the employed EAM potentials for geometrical
  and vibrational properties of metallic nanostructures is
  desirable. \emph{Ab initio} molecular dynamics could be used as a reference, and a
  survey of different interatomic potentials and force fields would be ideal,
  along the lines recently presented for bulk materials
  \cite{Choudhary2017}. Further extensions constitute the analysis of
  additional starting geometries or the temperature dependence of electronic
  and phononic thermal conductance contributions.

\section*{\label{sec:acknowledgements}Acknowledgements}
We thank J.\ Nieswand for preliminary studies during his Bachelor thesis
regarding ways to determine the phonon thermal conductance with
molecular-dynamics methods \cite{nieswand2016} and J.\ C.\ Kl\"ockner for
stimulating discussions on the comparison between molecular-dynamics and
density-functional-theory results for phonon heat transport.  D.O.M., M.M.,
P.N.\, and F.P.\ were funded by the Collaborative Research Center (SFB) 767 of
the German Research Foundation (DFG). F.M., M.M.\, and P.N.\ gratefully
acknowledge the Gauss Centre for Supercomputing e.V.\ (www.gauss-centre.eu)
for providing computing time through the John von Neumann Institute for
Computing (NIC) on the Supercomputer JUWELS at J\"ulich Supercomputing Centre
(JSC). An important part of the numerical modeling was furthermore carried out
on the computational resources of the bwHPC program, namely the bwUniCluster
and the JUSTUS HPC facility.

%\appendix

\section*{\label{sec:FullPull-ComputationalCost}APPENDIX: SINGLE GOLD STRETCHING PROCESS
  AND COMPUTATIONAL COSTS}

In this appendix, we present a full junction stretching event for Au. In
  addition, we discuss the computational costs of our simulations.

\begin{figure}
  \includegraphics[width=0.5\textwidth]{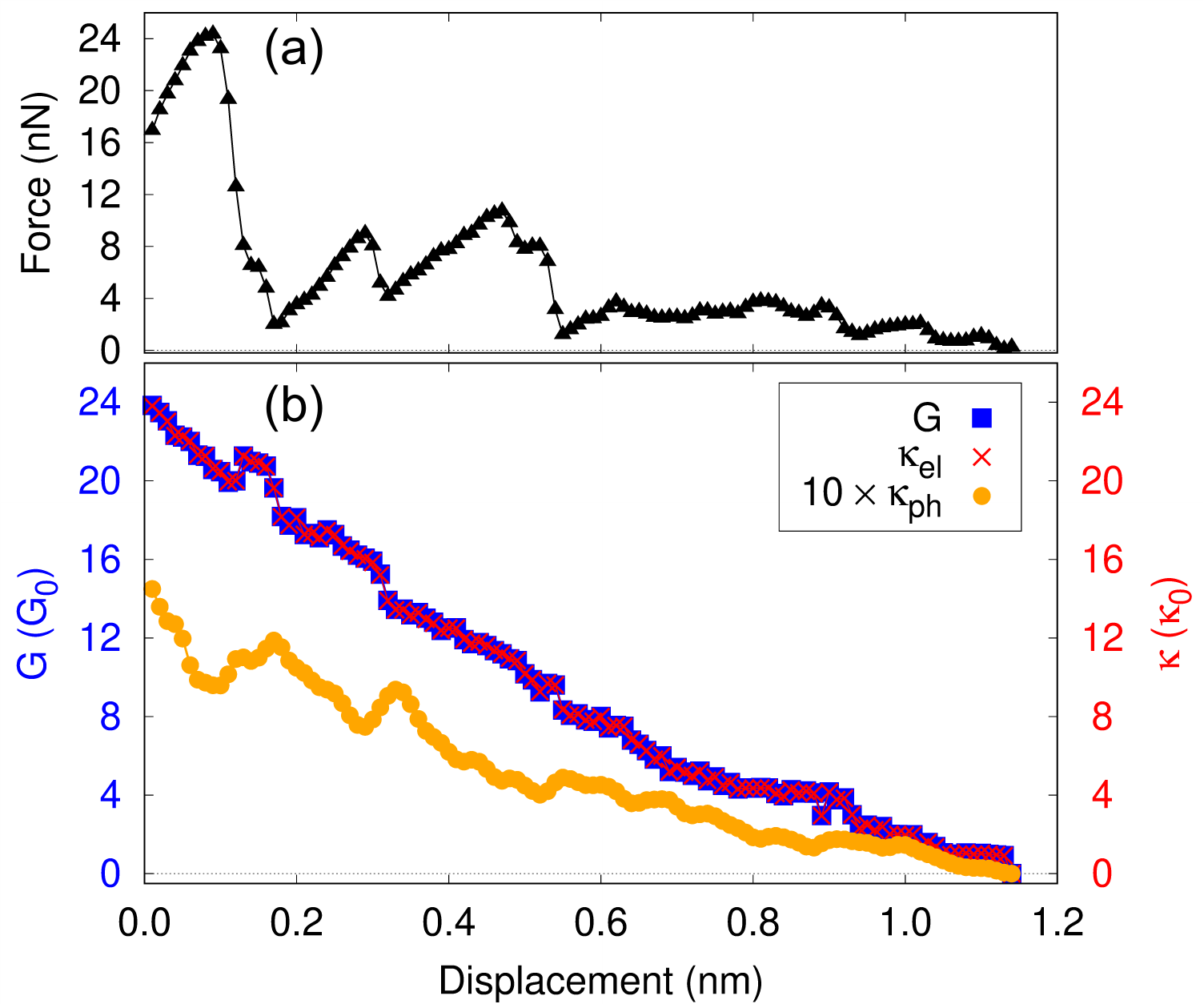}
  \caption{The single stretching process for Au of Fig.~\ref{fig:single_gold},
    but showing the full range of displacements analyzed.}
  \label{fig:single_gold_full}
\end{figure}

In Figs.~\ref{fig:single_gold}, \ref{fig:single_platinum}, and
  \ref{fig:single_aluminium} we showed a part of the full stretching
  processes, respectively, focusing on the end before contact rupture. For
  completeness we display in Fig.~\ref{fig:single_gold_full} the full
  stretching simulation for the Au junction of Fig.~\ref{fig:single_gold},
  i.e.\, we extend the plot to the full range of displacements studied.

In Fig.~\ref{fig:single_gold_full} it is nicely visible that $G$ and
  $\kappa_{\text{el}}$ follow each other closely for all separations, as
  expected from the Wiedemann-Franz law. It is also evident that these
  electronic transport properties exhibit jumps, when the force changes
  abruptly in plastic stages and atoms rearrange. During elastic stages $G$
  and $\kappa_{\text{el}}$ tend to decrease rather continuously as the
  nanowire narrows, but before rupture several rather horizontal conductance
  plateaus are found. We observe that $\kappa_{\text{ph}}$ decreases more or
  less continuously in elastic stages. But it typically increases during
  plastic stages. This is particularly pronounced at the beginning of the
  stretching process for displacements between 0 and 0.6~nm. An explanation
  for the behavior in terms of elastic constants has been given in the main
  text.

Let us now discuss computational requirements. As mentioned in the
  discussion of Fig.~\ref{fig:single_gold}, the contact breaks at a
  displacement of 1.16~nm. At the stretching speed of 0.01~m/s, a time step of
  1~fs and adding 3~ns or 3,000,000 simulation steps for the initial
  temperature increase and equilibrations (see
  Sec.~\ref{subsec:Geometries}), this corresponds to a simulation time of
  119~ns or 119,000,000 necessary molecular-dynamics steps.

Since the point of rupture is not known in advance, we actually computed
  203,000,000 molecular-dynamics steps for the simulation shown in
  Fig.~\ref{fig:single_gold_full}. For the NEMD-related part, including
  generation of geometries, force determination and phonon thermal transport
  calculations, we needed 146~core-h on an Intel Broadwell/Xeon E5-2680V4
  architecture. The separate electronic structure simulations for 120 atomic
  configurations required additional 935~core-h on an Intel Sandy Bridge/Xeon
  E5-2670 computer.

Let us now estimate the total computing costs required for this
  study. Taking into account that nanowires break at different time points, we
  need to multiply the computing time for gold roughly by 60, since this is
  the number of stretching processes analyzed. For Al the simulation effort
  is comparable to Au. Due to the relatively short lattice constant of Pt and
  the larger cutoff of the EAM potential as well as more iterations spent in
  the self-consistent loop required to achieve charge neutrality in the
  electronic transport simulations, the computational effort turns out to be a
  factor of 1.5 larger than for Au. Overall this means that we need to
  multiply the computing time above by a factor of 210, resulting in
  approximately 222,000~core-h for the whole study.

\providecommand{\noopsort}[1]{} \providecommand{\singleletter}[1]{#1}

\end{document}